\documentclass[12pt]{iopart}
\usepackage{graphicx}
\graphicspath{{graphics/}}
\usepackage[english]{babel}
\usepackage[utf8]{inputenc}

\expandafter\let\csname equation*\endcsname\relax
\expandafter\let\csname endequation*\endcsname\relax
\usepackage{amsmath}
\usepackage{amssymb}	
\usepackage[color=yellow!60]{todonotes}
\usepackage[retainorgcmds]{IEEEtrantools} 

\begin{document}
\title[Tunable condensate shape and scaling laws in pair-factorized
steady states]{Numerical survey of the tunable condensate shape and
  scaling laws in pair-factorized steady
  states}
\author{Eugen Ehrenpreis, Hannes Nagel and Wolfhard Janke}
\address{Institut f\"ur Theoretische Physik, Universit\"at Leipzig,
  Postfach 100~920, 04009~Leipzig, Germany}
\eads{
  \mailto{Eugen.Ehrenpreis@itp.uni-leipzig.de},
  \mailto{Hannes.Nagel@itp.uni-leipzig.de},
  \mailto{Wolfhard.Janke@itp.uni-leipzig.de} }

\begin{abstract}
  We numerically survey predictions on the shapes and scaling laws of
  particle condensates that emerge as a result of spontaneous symmetry
  breaking in pair-factorized steady states of a stochastic transport
  process.
  The specific model consists of indistinguishable particles that
  stochastically hop between sites controlled by a tunable potential.
  We identify the different condensate shapes within their respective
  parameter regimes as well as determine precisely the condensate width
  scaling.
\end{abstract}

\pacs{5.70.Fh,5.40.-a,5.70.Ln}
\maketitle

\section{Introduction}
In stochastic mass transport processes, one is generally concerned
with the dynamics of abstract, usually indistinguishable
particles. Well known models include the totally asymmetric exclusion
process
(TASEP)~\cite{Derrida92:Exact-TASEP,Schuetz93:Phase-Transitions-ASEP}
where the unidirectional free movement of particles is hindered by
other particles, and the more generic zero-range process
(ZRP)~\cite{Spitzer70:interaction_markov} that only models zero-ranged
interactions of particles.
Depending on the situation and given type of dynamics these particles
may represent actual molecules on the microscopic scale, mesoscopic
objects like intracellular motor
proteins~\cite{Chowdhury05:Transport-Traffic-Biology} or even
macroscopic bodies such as people or cars in traffic
systems~\cite{Chowdhury00:SP-Traffic}.
Interesting effects such as phase transitions in the one-dimensional
system and real-space condensation of particles can already be seen in
the ZRP, despite the absence of ranged
interactions~\cite{Evans00:phase-transitions-1d,Evans04:Spontaneous-Symmetry-breaking,Evans05:nonequilibrium-mechanics,Majumdar09:real-space-condensation-review}.
In this paper we will consider a generalized transport process
including short-ranged interactions which allows for a pair-factorized
steady state (PFSS)~\cite{Evans06:PFSS}. For a specific type of
interactions proposed by Wac{\l}aw et~al.\
\cite{Waclaw09:tuning-shape}, this model features a spatially extended
condensate that can assume several distinguished shapes. That is, by
parametrization the condensate can be tuned to a single-site peak,
rectangular or smooth parabolic envelope shape. Here we will consider
and validate predictions on the scaling of the condensate width as
well as its shape made in Refs.~\cite{Waclaw09:tuning-shape,Waclaw09:mass-condensation-1d}.

The rest of the paper is organized as follows. In the next section we
will introduce the considered stochastic transport model and summarize
the derivation of the predictions. The methods needed to simulate the
model and measure its properties will be discussed in the third
section followed by a discussion of our results in the fourth
section. The paper closes with our conclusions in the fifth section.

\section{Model}

Consider $M$ indistinguishable particles on a one-dimensional,
periodic lattice with $N$ sites. Any site $i$ can be occupied by any
number of particles $m_{i}\ge 0$. We only consider closed systems,
that is, the number of particles is conserved,
$M=\sum_{i=0}^{N}m_{i}$, with an overall density $\rho=M/N$.  The
dynamics is defined as a Markovian stochastic process: At any time
step, a particle may leave from a randomly selected site and move to
either of its neighbours.
The specific dynamics is largely controlled by the hopping rate
$u(m_{i}\vert m_{i-1},m_{i+1})$, which determines whether a particle
actually performs a hop. The fact, that it depends only on the
occupation numbers of the selected site and those of its direct
neighbours, reflects the short interaction range of particles and
sites.

The model can easily be tuned from symmetric to totally asymmetric
dynamics by introducing a probability $r$, that is used to decide
whether a particle hops to the right or left neighbour. Even though
this allows tuning the process from equilibrium ($r=1/2$) to
non-equilibrium ($r \neq 1/2$) dynamics, the stationary state remains
unaffected.

\subsection{Definitions}
Such a model was proposed by Evans et al.~\cite{Evans06:PFSS} as an
extension of the well known ZRP and therefore inherits several of its
properties: The model features a stationary state that may contain a
particle condensate. The steady state probabilities $P(\vec{m})$
factorize over symmetric non-negative weight functions $g(m,n)$ of
\emph{pairs} of sites
\begin{equation}
  \label{eq:generalProbability}
  P(\vec{m}) = P(m_{1},\ldots,m_{N}) = \frac{1}{Z} \prod_{i=1}^{N} g(m_{i},m_{i+1})
\end{equation}
for $\sum m_{i}=M$ kept constant,
with the partition function $Z = \sum_{\{\vec{m}\}} P(\vec{m})$ and
the configuration space $\{\vec{m}\}$. In contrast, for the ZRP, the
state probabilities factorize over \emph{single-site} weight functions.

If the weight function $g(m_i,m_i+1)$ falls off faster than any power
law, it has been shown that there exists a critical density
$\rho_\mathrm{c}$, above which translational symmetry is broken and a
particle condensate
emerges~\cite{Evans06:PFSS,Waclaw09:tuning-shape,Waclaw09:mass-condensation-1d}. That
is, in the steady state the bulk system can only hold a limited number
of particles given by $\rho_{\mathrm{c}}$ and any further added
particle increases only the condensate mass $M_{\mathrm{c}} =
M-\rho_{\text{c}}N = (\rho-\rho_{\text{c}})N$.

The hopping rate that leads to the steady
state~\eqref{eq:generalProbability} is easily determined with the weight
function $g(m,n)$ as
\begin{equation}
 u(m_{i} \vert m_{i-1},m_{i+1}) = 
 \frac{g(m_{i}-1, m_{i-1})}{g(m_{i},m_{i-1})}
 \frac{g(m_{i}-1, m_{i+1})}{g(m_{i},m_{i+1})}
\end{equation}
by resolving the balance condition in the steady state. In the case of
symmetric hopping, detailed balance is fulfilled.

An interesting way to construct such weights is to simplify and separate
\begin{equation}
    \label{eq:generalWeights}
 g(m,n) = \sqrt{p(m)p(n)} K(\left\vert m-n \right\vert),
\end{equation}
i.e., to factorize $g(m,n)$ into a zero-range part $p(m)$ and a
short-range interaction part $K(\vert m-n \vert)$ that depends on the
difference of nearest-neighbour occupation numbers. The term
\emph{zero-range} interaction refers to the fact that it allows a
particle on a given site to interact with particles on the given site
only. In contrast, \emph{short-range} interactions act between
particles of different sites. For example, the zero-range
process~\cite{Spitzer70:interaction_markov} with the hopping rate
$u(m)=1+b/m$ is easily implemented using the weight function
$g(m,n)=\frac{1+b/m}{b+1}=p(m)$, with $K(x)=1$ as there is no
short-range interaction involved.

In the original model by Evans et~al.\ \cite{Evans06:PFSS} governed by
the weights
\begin{equation}
  \label{eq:weights-evans}
  K(x)=\exp\left( -J \vert x \vert \right),\;
  p(m)=\exp\left( U_{0} \delta_{m,0} \right),
\end{equation}
the condensate has a characteristic smooth parabolic shape. Under the
conditions that $K(x)$ falls off faster than any power law and
$p(m)=\mathrm{const}$ for some $m>m_{\text{max}}$, the exact envelope
shape can be calculated as well as its fluctuations and the condensate
width scaling behaviour
$W\propto\sqrt{M_{\text{c}}}$~\cite{Waclaw09:mass-condensation-1d}.
The factorization \eqref{eq:generalWeights} of $g(m,n)$ also makes it
easier to understand the physical picture of condensation
involved. The zero-range potential $p(m)$ gives a penalty to
increasing occupation numbers, much like a potential energy term,
while the short-range interaction $K(x)$ tends to reduce the
difference in occupation numbers of neighbouring sites, acting like a
surface energy. It is then the respective relative importance of these
terms that governs the properties of the emerging condensate.

Wac{\l}aw et~al.~\cite{Waclaw09:tuning-shape} then suggested a family
of weights
\begin{equation}
	\label{eq:WaclawWeights}
	K(x) \sim \e^{-a\vert x\vert^{\beta}},\quad
	p(m) \sim \e^{-bm^{\gamma}}
\end{equation}
that allows one to deliberately violate the asymptotic condition on
$p(m)$ for large $m$ by choice of parameters. This produces regimes
featuring qualitatively different condensation properties, and most
notably may be used to tune the condensates envelope shape as well as
the scaling behaviour of its width with system
size~\cite{Waclaw09:tuning-shape,Waclaw09:mass-condensation-1d}. In
Ref.~\cite{Waclaw09:mass-condensation-1d} these regimes and properties
are derived in the large-volume limit of the model. It is the subject
of this work to check these predictions by means of numerical
simulations at finite volumes and scaling analyses. In the next
subsection we will therefore shortly summarize the derivation of the
interesting new properties.

\subsection{Theoretical predictions}

The proposed model is fully defined by
Eqs.~\eqref{eq:generalProbability},~\eqref{eq:generalWeights},
and~\eqref{eq:WaclawWeights}. The parameters $\beta,\gamma$ give the
respective growth behaviour of the short-range and zero-range
interaction strengths and are expected to have the most influence on
condensation properties. The factors $a,b$ allow to tune the relative
strengths of the interactions but do not change the qualitative
behaviour of the system as we will see later. In this paper we will
therefore use $a=b=1$.

First, the condition $\gamma \le 1$ is derived to observe condensation
at all. Next, the steady state weight
Eq.~\eqref{eq:generalProbability} can be thought to factor into one
part accounting for the statistical weight of the condensate and one
for that of the rest of the system. Then, a comparison of the weight
of a single-site condensate $P_{1}$ and that of a condensate occupying
two or more sites $P_{n}$ shows that for $\beta>\gamma$ the weight
$P_{n}$ grows faster than $P_{1}$, so that an extended condensate can
be expected in the large system-size limit.

\begin{figure}
  \centering
  \includegraphics{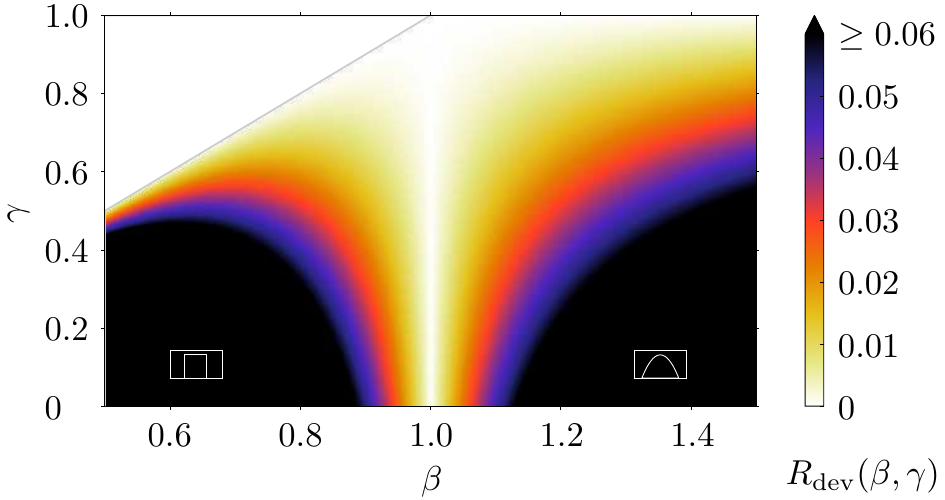}
  \caption{Comparing the scaling behaviours of the most likely
    rectangular and smooth condensate shapes using the relative
    deviation (\ref{eq:relative_deviation}), over a region in
    $\beta$-$\gamma$-parameter space. Larger deviations $R_{\mathrm{dev}}$ (cut off at
    $0.06$) mean that (in the thermodynamic limit) the probabilities
    are different, and lower ones that they are close.}
  \label{fig:rectSmoothDifferentiationPlot}
\end{figure}

\begin{figure}
  \centering
  \includegraphics{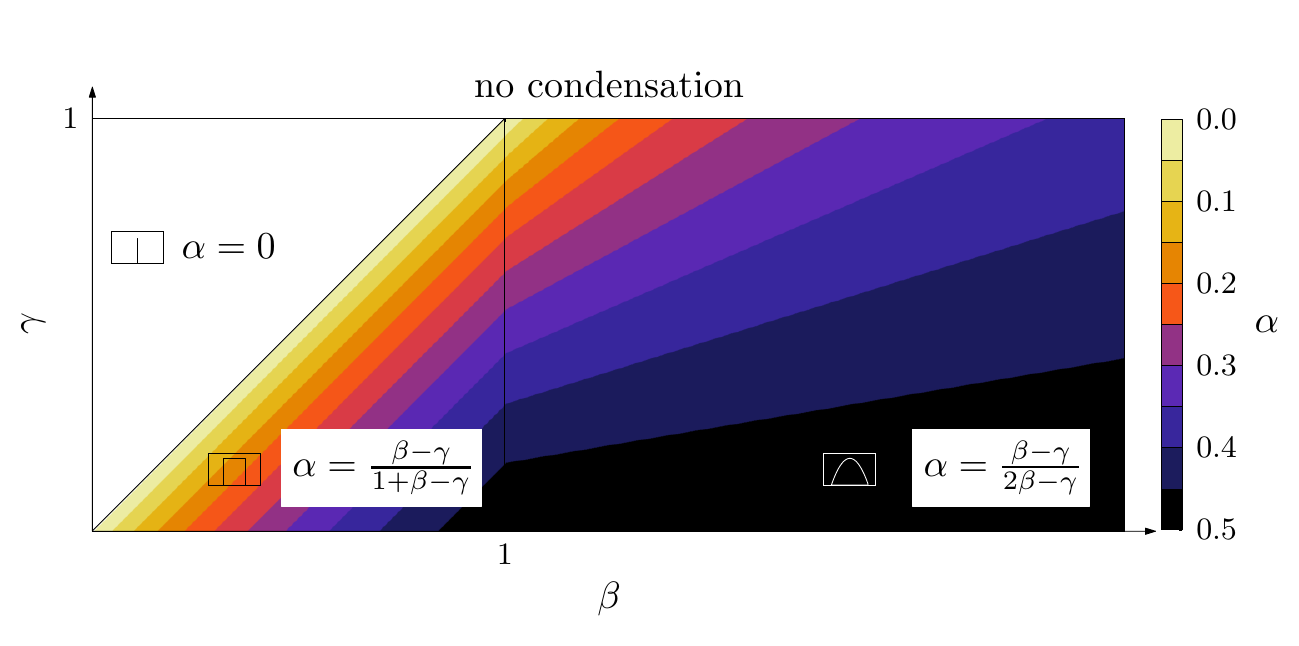}
  \caption{The width-scaling exponent $\alpha$ in $W\propto
    M_{\mathrm{c}}^{\alpha}$ for different parametrizations $\beta$ and
    $\gamma$, cf.~Eq.~\eqref{eq:widthScaling}.  The three
    regions of different scaling behaviour mark the
    distinctive phases. For $\gamma\ge 1$ there is no
    condensation.  This figure is a reproduction of Fig.~5
    in~\cite{Waclaw09:mass-condensation-1d}.}
  \label{fig:phaseDiagramShapeAdjustingWeights}
\end{figure}

In the case of the extended condensate a similar approach is used to
predict some of its properties. Under the assumption, that the
contribution of fluctuation of occupation numbers is small compared to
the contribution from the condensate itself, the weight of a
condensate extended over $W$ sites is given as~\cite{Waclaw09:mass-condensation-1d}
\begin{equation}
\label{eq:fixed-envelope-weight}
\ln P(W) \approx
Wc + \sum_{i} \ln K\left( \langle m_{i+1} - m_{i} \rangle \right)
+ \sum_{i} \ln p\left( \langle m_{i} \rangle \right),
\end{equation}
where for simplicity $c=s-\ln \lambda_{\mathrm{max}}$ is treated as a
positive constant to approximate the partition function, with an
entropic term $s$ that corresponds to fluctuations of the occupation
numbers and the largest eigenvalue $\lambda_{\mathrm{max}}$ of the
associated transfer matrix.
The expectation values $\langle m_{i} \rangle$ correspond to values
$Hh(2i/W-1)$ where $h(\xi)$ with $\xi=2i/W-1$ is the rescaled envelope
shape of a condensate with height $H=M_{\text{c}}/W$ and width
$W$. Now, for the expectation of occupation number differences
$\langle m_{i+1} - m_{i} \rangle$ a distinction into two cases is
proposed to allow simplifications. Either, the envelope shape $h(\xi)$
is smooth or it is of rectangular shape. In the first case the
expectation of the difference can be written using the derivative of
the shape $h'(\xi)$ and the expectation reads
$(2M_{\text{c}}/W^{2})h'(\xi)$. In the second suggested case, all
differences are zero except at the condensate boundaries and the sum
reduces to a single term. Then in both cases, the sums can be written
as integrals using the rescaled shapes resulting in
\begin{align}
  \ln P_{\text{smooth}}(W)\approx W \left[ c + \int_{0}^{1}\mathrm{d}\xi \ln K\left( \frac{2M_{\text{c}}}{W^{2}}h'(\xi) \right) + \int_{0}^{1}\mathrm{d}\xi \ln p(Hh(\xi)) \right], \\
  \ln P_{\text{rect}}(W)  \approx Wc + 2\ln K\left( h(1)\frac{M_{\text{c}}}{W} \right) + W \int_{0}^{1} \mathrm{d}\xi \ln p(Hh(\xi)).
\end{align}
The sums decouple into a prefactor and a constant integral term, which
does not depend on the condensate width. The prefactors however are used to
determine the growth of the different condensate weights depending on
their respective widths and masses
\begin{align}
  \ln P_{\text{smooth}}(W) \approx W \left[ c
    - a \left( \frac{2M_{\text{c}}}{W^{2}}\right)^{\beta} \int_{0}^{1}\mathrm{d}\xi \vert h'(\xi)\vert^{\beta}
    - b\left(\frac{M_{\text{c}}}{W}\right)^{\gamma} \int_{0}^{1}\mathrm{d}\xi \vert h(\xi)\vert^{\gamma}
    \right], \\
    \ln P_{\text{rect}}(W) \approx Wc - a \left( \frac{M_{\text{c}}}{W}\right)^{\beta} 2\vert h(1)\vert^{\beta}
    -bW\left( \frac{M_{\text{c}}}{W}\right)^{\gamma} \int_{0}^{1}\mathrm{d}\xi \vert h(\xi)\vert^{\gamma}.
\end{align}
Finally, the condensate weights are estimated\footnote{During the
  estimation of the maximal weight for the given condensate type
  additional constraints are
  found~\cite{Waclaw09:mass-condensation-1d}, where no solution
  exists, narrowing down the number of cases to look at.}  by finding
their maximal value with respect to the condensate width (with
constant mass) which also gives the scaling laws for the condensate
width
\begin{align}
  \label{eq:derivation_lnP_smooth}
  &\ln P_{\text{smooth}}(W) \sim -M_{\text{c}}^{(\gamma\beta+\beta-\gamma)/(2\beta-\gamma)}, \quad
  W \sim M_{\text{c}}^{(\beta-\gamma)/(2\beta - \gamma)} \;\text{for}\; \beta>\frac{1}{2},\, \text{and}
  \\
  \label{eq:derivation_lnP_rect}
  &\ln P_{\text{rect}}(W) \sim -M_{\text{c}}^{\beta/(\beta-\gamma+1)}, \quad
  W \sim M_{\text{c}}^{(\beta-\gamma)/(\beta-\gamma+1)} \;\text{for}\; \beta>0
\end{align}
in the smooth and rectangular case, respectively. For $0<\beta<1/2$,
no solution exists for smooth condensates. Finally these weights are
compared to find $\ln P_{\text{rect}}(W) >\ln P_{\text{smooth}}$ for
$1/2<\beta<1$, showing that the rectangular shape is the stable one in
this regime. In Fig.~\ref{fig:rectSmoothDifferentiationPlot} we plot
the ($M_{\text{c}}$-independent) relative deviation
\begin{equation}
\label{eq:relative_deviation}
R_{\text{dev}}(\beta, \gamma) = \frac{\log (-\log P_{\text{smooth}}) - \log (-\log P_{\text{rect}})}{\log (-\log P_{\text{rect}}) }
\end{equation}
to give an impression of the regime
boundary. Figure~\ref{fig:phaseDiagramShapeAdjustingWeights} shows
these regimes and the respective condensate scaling laws.

\begin{equation}
  \label{eq:widthScaling}
  \alpha_{\text{peak}} = 0\,,
  \quad
  \alpha_{\text{smooth}} = \frac{\beta-\gamma}{2\beta-\gamma}\,,
  \quad
  \alpha_{\text{rect}} = \frac{\beta-\gamma}{\beta-\gamma+1}\,.
\end{equation}

\section{Methods}

We will employ the empirical method of taking samples while evolving a
Markov process. Once we have obtained a set of representative samples
for one particular parametrization ($\beta$, $\gamma$,
$M_\mathrm{c}$), we can determine the condensate width. With the
widths for different condensate volumes $M_\mathrm{c}$, we then can
estimate the condensate-width scaling exponent $\alpha$.

\subsection{Generating system samples}
The straightforward method for generating samples is to simulate the
process as defined. Its dynamics with typically small transition
probabilities leads to a large waiting time until the gathered samples
are representative. The key ingredient for improving the simulation is
that one knows the steady state probabilities 
(\ref{eq:generalProbability}) which are invariant under the employed
dynamics, asymmetric hopping ($r \neq 1/2$) in non-equilibrium or
symmetric hopping ($r=1/2$) in equilibrium. In the latter case, we can
verify stationary equivalence to systems with other dynamics that
yield the same distribution, but mix much more quickly. The Metropolis
dynamics is known to mix well and can be readily adjusted for the
desired distribution, which is why we chose to use it in our
simulations. When the update scheme is altered so that particles may
hop to any site instead of just the nearest neighbours, the simulation
algorithm assumes the form:
\begin{enumerate}
\item Select randomly a site from which a move of one particle is
  attempted.
\item If this site is empty, the system remains in the same state for
  another time step of $1/N$ sweeps. If it is occupied, choose any of
  the other sites as a destination site.
\item Accept the move with Metropolis probability $P_{\text{acc}} :=
  \min{(P_{\text{final}}/P_{\text{initial}}, 1)}$ or remain at the
  current state for another time step otherwise, where
  $P_{\text{initial}}$ and $P_{\text{final}}$ are the probabilities of
  the microstates before and after the proposed move, respectively.
\end{enumerate}
For our implementation of the simulation we use the reference
implementation of a variant of the Mersenne Twister pseudo-random
number generator~\cite{Matsumoto:1998MT,Saito:2009vz}.

\begin{figure}
  \centering
  \includegraphics{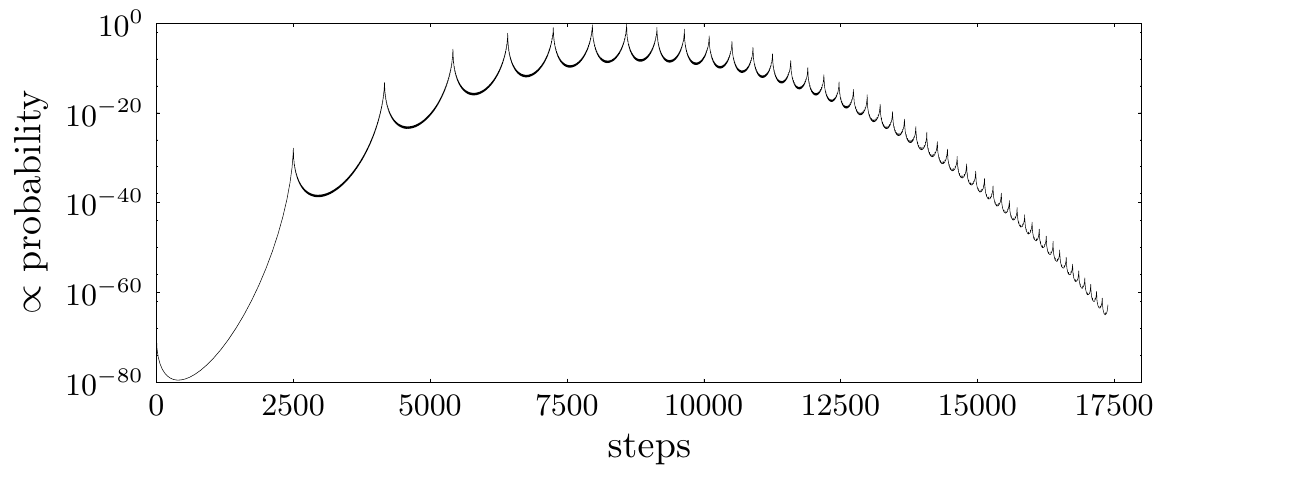}
  \caption{The shortest path in probability landscape starting from a
    condensate occupying a single site with 5000 particles and going
    to rectangular condensates of increasing widths. The peaks
    correspond to the high statistical weights of rectangular
    condensates and the valleys to those of the transitional states on
    a an ideal trajectory. The (unnormalized) probabilities are based
    on the parametrization with $\beta=0.6$, $\gamma=0.4$.}
  \label{fig:shortestPath}
\end{figure}

This approach works well for the smooth condensate shape region
($\beta \ge 1$), where the probability landscape is marginally
rugged. For systems in the rectangular condensate regime this is not
the case anymore because the probability landscape is heavily rugged
as shown in Fig.~\ref{fig:shortestPath}. The transitional states
between rectangular condensates of different widths are suppressed by
several orders of magnitude and simulations with single-particle (or
local) updates mix slowly.

In order to address this issue, we designed an additional set of
multi-particle (non-local) updates that allow jumps between high
probability regions, which correspond to the peaks in
Fig.~\ref{fig:shortestPath}, in a single step, thus bypassing the
suppressed transitional states. This is achieved by defining a
condensate expansion and reduction transformation, where the
condensate width is increased or decreased by about one unit.

With the above stated local update algorithm it is easy to verify that
the resulting stationary distribution is given by
Eqs.~\eqref{eq:generalProbability},~\eqref{eq:generalWeights},~\eqref{eq:WaclawWeights}. By
changing the dynamics we must be careful to keep the steady state
unchanged. The local update performs well in exploring the landscape
locally, which is why we will keep it and independently add non-local
updates. The probability currents added by the new updates must cancel
each other out at every state. This can be achieved by constructing a
dynamics fulfilling detailed balance, where an incoming probability
current is being canceled by its corresponding outgoing one. The one
we used before and which we will use again is based on Metropolis
transition rates.
The main problem here is to treat equally the two non-local update
moves and how often to use them in order to minimize the relaxation
time.
We chose to solve this by restricting the number of new transitions
that are permitted at any state to (at most) two; an expansion and a
corresponding reduction transition. The new simulation algorithm then
consists of a preceding step, where a decision (according to some
predefined probability) is made on whether to attempt a local or a
non-local step\footnote{For small probabilities it is better to use
  the waiting time, until a non-local update will be performed, which
  is geometrically distributed.}. Choosing a local update is followed
by the algorithm above, while a non-local update is governed by the
following algorithm:
\begin{enumerate}
\item Decide whether to attempt an expansion or reduction
  transformation with equal probability.
\item If the chosen transformation is not valid, the system remains at
  the current state for another time step (the system is constrained
  in a similar way to the situation of attempting to move a particle
  from an unoccupied site), otherwise continue.
\item Accept the transformation with Metropolis probability
  $P_{\text{acc}} = \min{(P_{\text{final}}/P_{\text{initial}}, 1)}$ or
  remain at the current state for another time step otherwise.
\end{enumerate}

\begin{figure}
  \centering
  \includegraphics[width=13.5966cm]{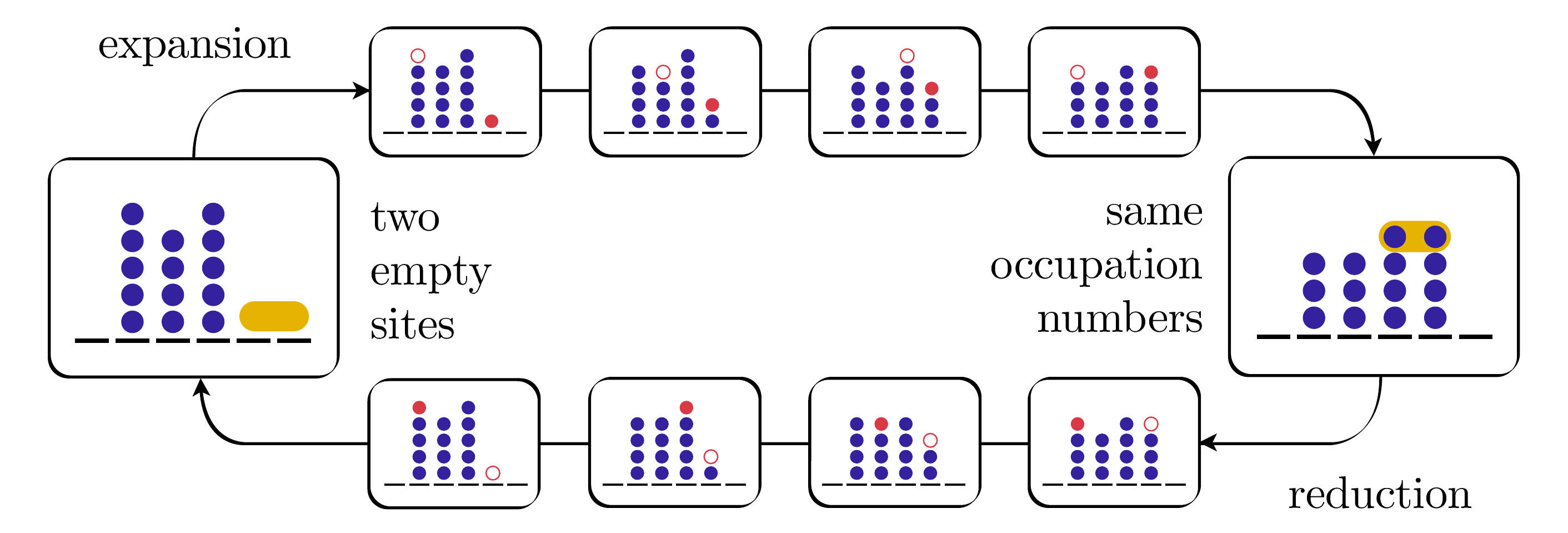}
  \caption{The transformation loop for an example condensate state
    including all intermediate steps. Any condensate state is suited
    for the respective transformation if it fits either starting point
    in this loop and adheres to the highlighted characteristics.}
  \label{fig:transformation loop}
\end{figure}

From the algorithm above, the necessary conditions are that there
should be only one way (if any) to expand or reduce the condensate and
the transformation must be invertible such that reducing a previously
expanded state always results in the original state.

The conditions we decided on is that expansion transformations are
eligible only if the original state has two unoccupied sites to the
right of the condensate. One removes particles from the top of the
condensate by sweeping from left to right and puts them on the empty
site right to the condensate, until the two last condensate sites have
the same occupation numbers (in some cases this is not possible, then
the transformation is invalid). Likewise, reduction transformations
are only valid if the original state has the same occupation numbers
at the two rightmost condensate sites. One removes the particles at
the rightmost condensate site and distributes them by sweeping from
left to right onto the rest of the
condensate. Figure~\ref{fig:transformation loop} gives an example of
both transformations.

For $M=1000$ particles with $\beta=0.7$ and $\gamma=0.4$ the
autocorrelation time of the condensate width of purely local updates
is about 10 times larger than that of non-local updates. Increasing
the particle volume by 50\% (100\%) increases this factor to 50
(250). Whereas the new method enables systems with more than $10^6$
particles to mix sufficiently in less than an hour, purely local
updates are constrained to less than $10^4$ particles.

In cases where $\beta$ and $\gamma$ are both close to $0$, the
critical density is so large, that the condensate is not surrounded by
unoccupied sites anymore. Our non-local updates are then rendered
ineffective and there is no easy way to adjust them to this situation.

\subsection{Determining the condensate width}

\begin{figure}
  \centering
  \includegraphics{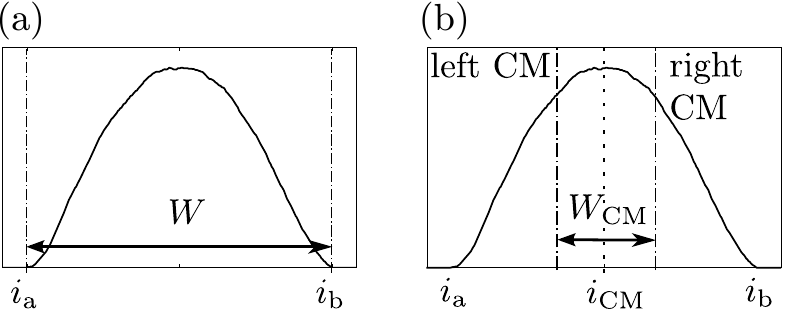}
  \caption{\label{fig:width measurement methods} Condensate width
    determination method by (a) taking the direct condensate base
    extension or (b) taking the distance of the left and right
    condensate wings respective centers of mass $W_{\text{CM}}$.}
\end{figure}

In the analytic approximation of the scaling
behaviour~\cite{Waclaw09:mass-condensation-1d}, the width of the
condensate is defined as the distance between the two outmost sites of
the condensate, which we will call the condensate extension $W$ (see
Fig.~\ref{fig:width measurement methods}(a)). Whereas this is a good
choice in the thermodynamic limit, for finite systems the condensate
edges are subject to interactions with the rest of the system, which
results in the edges being smeared outward the condensate. The
condensate extension is an observable that is strongly influenced by
these finite-size effects, and therefore overestimates the condensate
width. This is especially the case with narrow but high condensates
that have steep edges.

\begin{figure}
  \centering
  \includegraphics{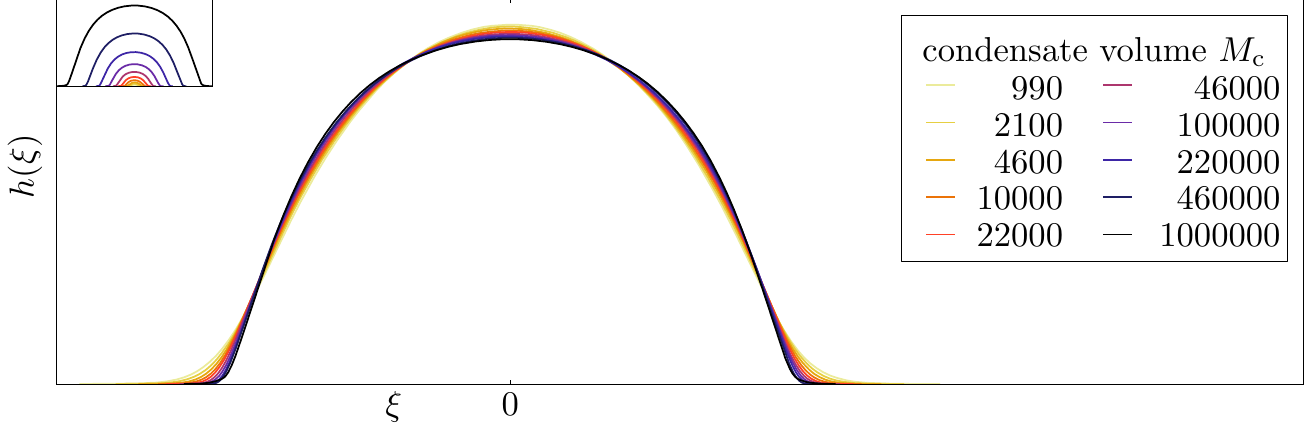}
  \caption{Comparison of mean condensate shapes at different volumes
    but the same parametrization $\beta=1.2$ and $\gamma=0.3$. The
    shapes were rescaled in terms of extension by the condensate width
    and in terms of occupation numbers by normalization to a unit
    volume. With increasing condensate volumes the shapes converge to
    a single shape which indicates a common \emph{characteristic
      shape} for this parametrization. The inset shows the original
    mean condensate shapes.}
  \label{fig_invariantCondensateShape}
\end{figure}

We decided to measure the condensate width with another observable
that is less prone to these smearing-out effects. This observable is
based on the assumption that given a parametrization $\beta$ and
$\gamma$, the system exhibits a common characteristic shape that is
merely rescaled in width and height in dependence on the volume of the
condensate. Reversing this assumption, it must be possible to rescale
condensate shapes of different volumes to such a common shape, which
was done for one example system in
Fig.~\ref{fig_invariantCondensateShape}. One can see there, that as
condensate volumes increase, the shapes converge to a characteristic
shape, with small volume condensates deviating from it due to the
mentioned finite-size effects.

First the condensate is separated in two wings left and right of its
center of mass (CM) position $i_{\text{CM}}=\sum_{i=i_{a}}^{i_{b}} i
m_{i} / M_{\text{c}}$, where $i_{a}$ and $i_{b}$ are the begin and end
positions of the condensate so that $m_i>0$ for $i_{a}<i<i_{b}$. Then
we calculate the width as the distance between the individual centers
of mass of these two wings
\begin{equation}
  \label{eq:WCM}
  W_{\text{CM}}=
   \frac{2}{M_{\text{c}}} \left( \sum_{i\ge i_{\text{CM}}}^{i_{b}}i m_{i} - \sum_{i_{a}}^{i<i_{\text{CM}}}i m_{i} \right).
\end{equation}
Figure~\ref{fig:width measurement methods} summarizes how this is done
in comparison to the direct estimation method.

\subsection{Determining the condensate width scaling behaviour}

\begin{figure}
  \centering
  \includegraphics{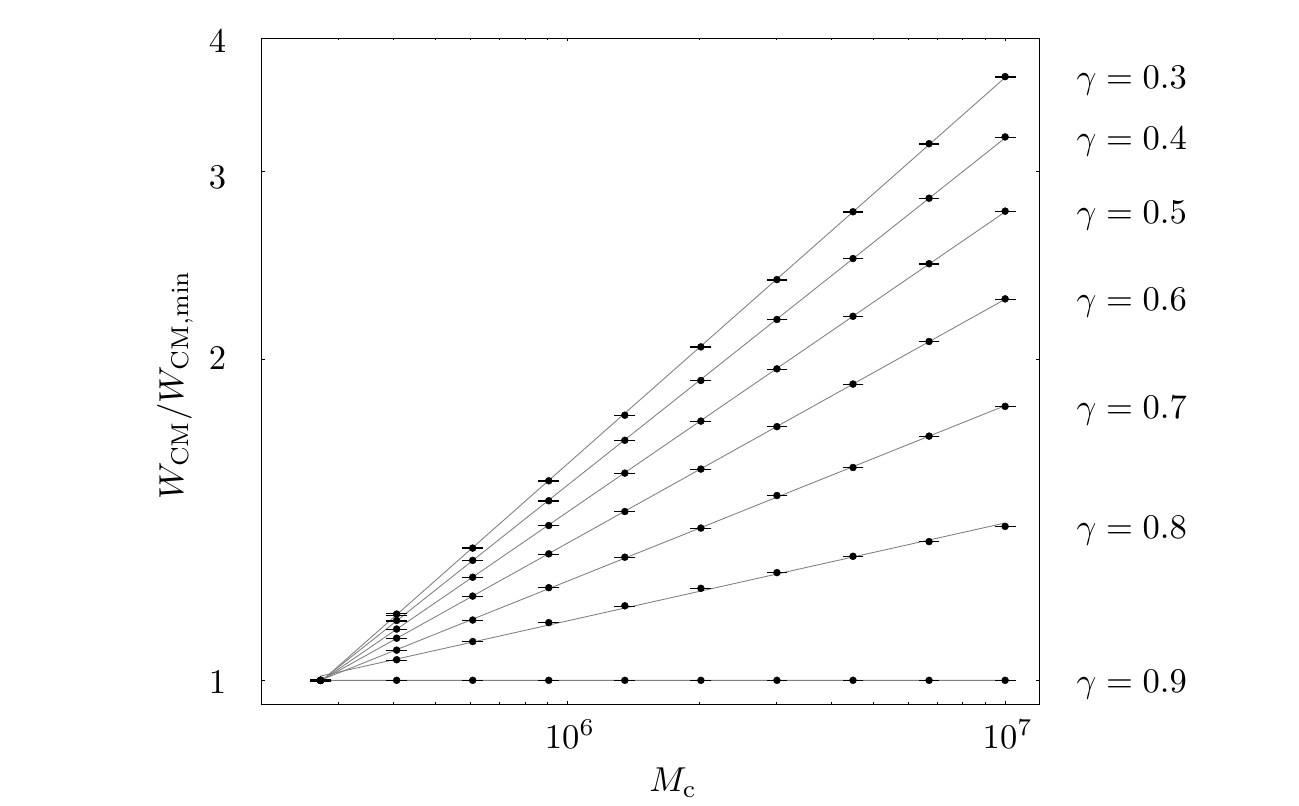}
  \caption{Log-log plot of the width scaling behaviour in dependence
    on the condensate volume $M_{\text{c}}$ for systems with
    $\beta=0.9$ and various $\gamma$ in the range $[0.3, 0.9]$. The
    wing center-of-mass widths (\ref{eq:WCM}) have been normalized by
    dividing them by the smallest width. }
  \label{fig:scalingLinearRegression}
\end{figure}

Once we have obtained estimates of the wing center-ov-mass width
observable $W_{\mathrm{CM}}$ for system parametrizations with
different condensate volumes $M_\mathrm{c}$, the
prediction~\cite{Waclaw09:mass-condensation-1d} states that for large
volumes the scaling behaviour adheres to $W_{\mathrm{CM}}\propto
M_\mathrm{c}^\alpha$ with $\alpha(\beta,\gamma)$. Taking the logarithm
this relation reduces to a linear relationship with $\alpha$
determining the slope. Figure \ref{fig:scalingLinearRegression}
confirms that the data asymptotically approaches this linear
relationship with increasing condensate volumes. Determining the
scaling exponent is a linear regression problem, where the measured
condensate width is uncertain but the measured condensate volume is
considered certain. A minor problem is the overestimation of the
condensate width for small condensates with strong smearing
(finite-size) effects, which we simply avoid by simulating large
enough condensates.

\begin{figure}
  \centering
  \includegraphics{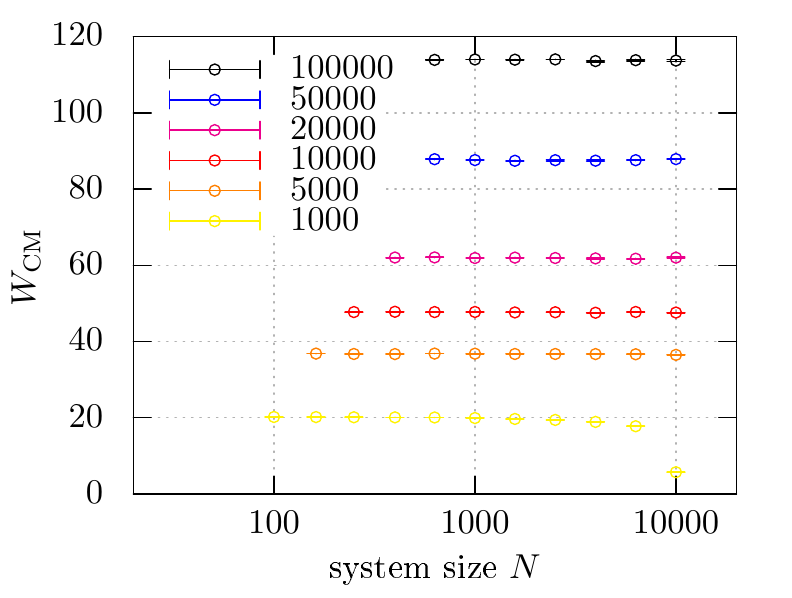}
  \caption{The dependence of the condensate width with a fixed
    condensate volume $M_\mathrm{c}$ on the system size for
    $\beta=1.3$ and $\gamma=0.5$. The critical density is about $\rho
    _\text{c} \simeq 0.14$.}
  \label{fig:shapeSystemSizeIndependence}
\end{figure}

One potentially problematic aspect left untouched so far, is how the
condensate width and with it the scaling behaviour is affected by the
system size, especially considering effective interactions between the
condensate edges through the background. The measurements in
Fig.~\ref{fig:shapeSystemSizeIndependence} show that for systems where
the condensate volume stays the same, but the system size changes, the
measured condensate width remains unchanged as long as $\rho >
\rho_{c}$. This leads to the conclusion that for densities much larger
than the critical density, the competition between the gaseous
background and condensate phase is dominated by the latter. This
system size independence is favorable for us, since it eliminates one
degree of freedom in the discussion (leaving only the system
parameterization $\beta$, $\gamma$ and condensate volume
$M_\mathrm{c}$) and allows us to choose systems sizes that are
sufficient for the condensates, but small enough to maintain
simulation performance. We used a fixed size about ten times the width
of the largest condensate to minimize interaction of condensate edges
around the boundary and eliminate percolating condensates.

\section{Results}
We are first going to look at the condensate phases, where especially
the boundary between the smooth and rectangular condensate phase is of
interest. Then we will present the estimated condensate width scaling
behaviour.

\subsection{Condensate phases}

\begin{figure}
  \centering
  \includegraphics{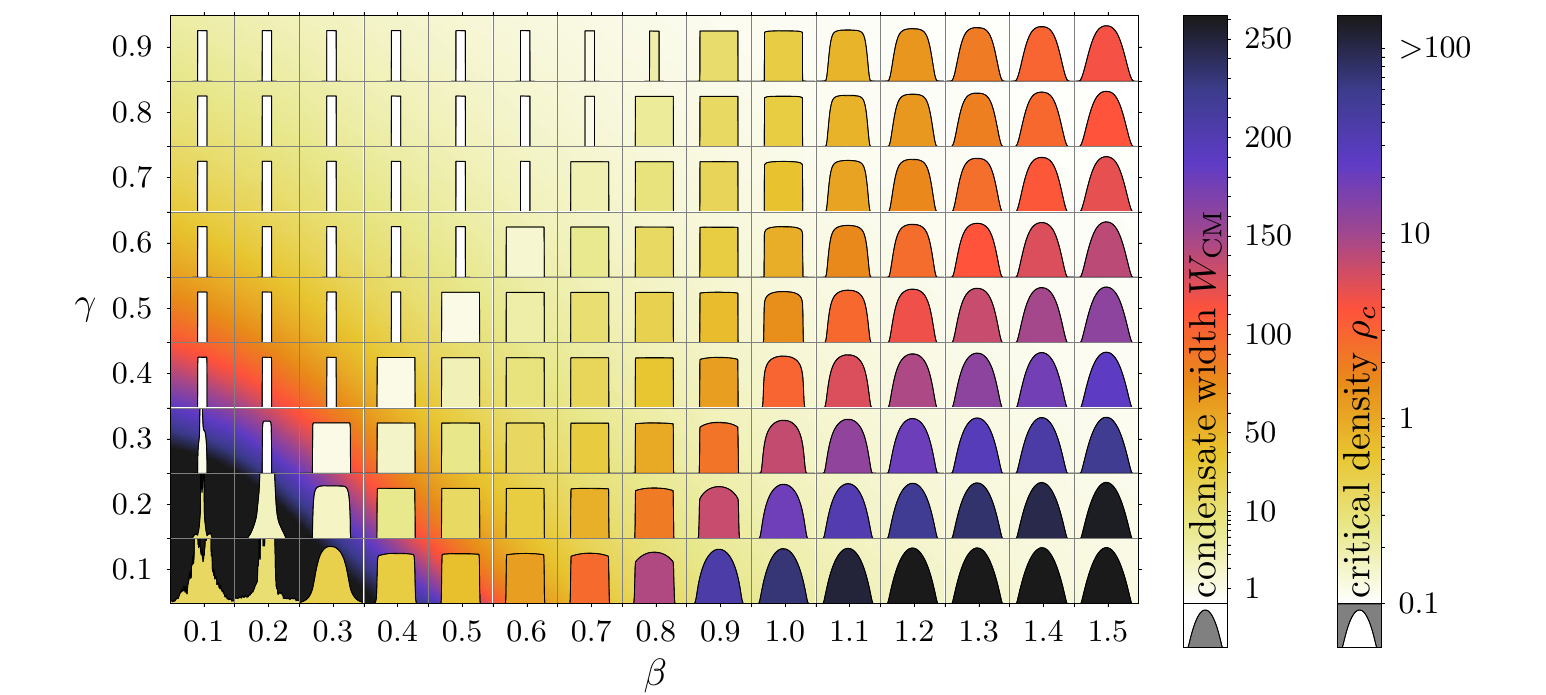}
  \caption{Comparison of the characteristic shapes for systems of
    various $\beta$ and $\gamma$ at a condensate volume of about
    $10^5$ masses. The shapes are formed by rescaling the width and
    height of all measured condensate sample shapes and only then
    averaging them (this avoids averaging artifacts one would get when
    interchanging these two steps). The fill colour inside the
    condensate shapes encodes the respective measured condensate width
    while the background colour around the shapes gives the critical
    density of the system. The shapes in the single-site condensate
    regime are plotted narrowed to give better distinction to extended
    shapes.}
  \label{fig:condensateGrid}
\end{figure}

In order to determine the phases we will use the previously introduced
concept of characteristic shapes. Figure~\ref{fig:condensateGrid}
shows estimations of these shapes for a range of points in parameter
space at a system volume of about $10^{5}$ masses. 
The data is obtained in runs with on the order of $10^{8}$ Monte Carlo
sweeps (consisting of $N$ single-particle moves) taking $10^{4}$
samples of the system.
For $\beta\lesssim 1$ we see characteristic shapes with constant
elevation in the middle section and steep edges at the boundaries,
which correspond to the predicted rectangular condensate shape. For
small $\beta$ and $\gamma$ we see that the estimation of the
characteristic shape fails. This is due to the increased critical
density and the associated failure of the non-local update. The
detection method used here is unable to distinguish between extended
rectangular condensates and peak-shaped single-site condensates.

\begin{figure}
  \centering
  \includegraphics{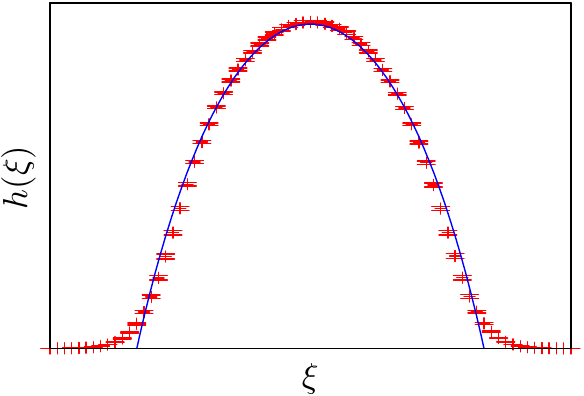}
  \caption{Comparison of the rescaled condensate shape for $\beta=1,
    \gamma=0.2, M=1000$ with the exact
    shape~(\ref{eq:pfss-envelope-shape}) derived in
    Ref.~\cite{Waclaw09:mass-condensation-1d} (blue line) for the
    interaction driven process~\eqref{eq:weights-evans}.}
  \label{fig:compare-pfss-shape}
\end{figure}

For $\beta\gtrsim 1$ we see characteristic shapes with decreasingly
steep edges that converge to a dome. These bell-shaped condensates
correspond to the smooth case that was predicted for this region. For
small $\gamma$ and $\beta\approx 1$ the condensation process is close
to the purely interaction driven process~\eqref{eq:weights-evans} for
which the characteristic shape
\begin{equation}
  h(\xi)=\frac{w}{2v}\log\left[
    \frac{ \cosh J - \cosh v\xi }{ \cosh J - \cosh v }
  \right], \; \text{with}\, \xi=\frac{2i}{W}-1, w=2.2005, v=0.5413, J=1 
  \label{eq:pfss-envelope-shape}
\end{equation}
can be analytically approximated~\cite{Waclaw09:mass-condensation-1d},
as compared in Fig.~\ref{fig:compare-pfss-shape}. The measured shape
is similar to this approximation and as $\beta$ increases, this
characteristic shape seems to describe large $\gamma$ systems as well.

\begin{figure}
  \includegraphics[clip]{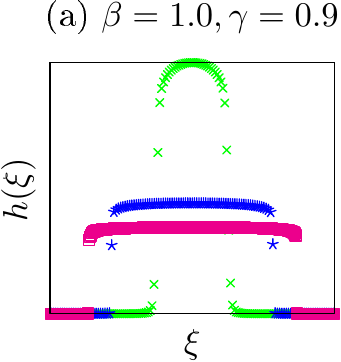}\hfill
  \includegraphics[clip]{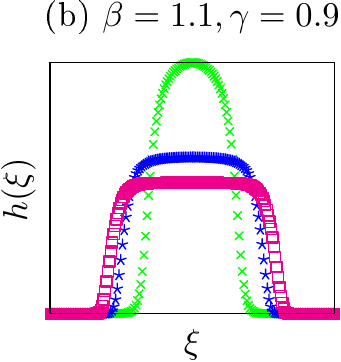}\hfill
  \includegraphics[clip]{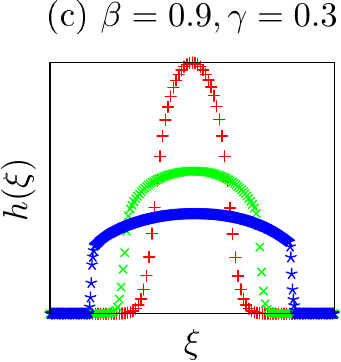}\hfill
  \includegraphics[clip]{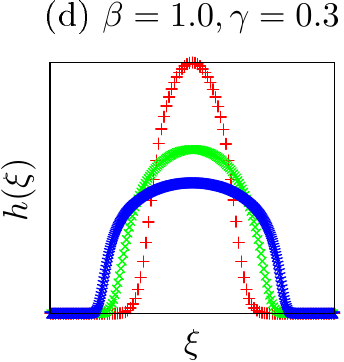}\hfill
  \includegraphics[clip]{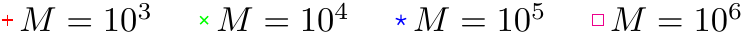}
  \caption{Finite-size effects of the condensate shape near the
    rectangular/smooth transition. The total amounts of masses in the
    system corresponding to the plotted rescaled shapes are with
    increasing width $M=10^4$, $M=10^5$, $M=10^{6}$ for $\gamma=0.9$
    and $M=10^{3}$, $M=10^{4}$ $M=10^{5}$ for $\gamma=0.3$
    respectively. For $\beta<1$, the smooth condensate edges disappear
    for increasing particle numbers $M$ while they stay for
    $\beta>1$.}
  \label{fig:transition-fss}
\end{figure}

The phase boundary in our measurements is not a clear separation at
$\beta=1$ between the two phases, as the predictions suggest, but it
seems to be smeared out in favour of smooth condensates for small
$\gamma$ and in favour of rectangular condensates for large
$\gamma$. In the respective regions the measured shapes combine
characteristics of the individual phases. Once we have steep edges
with smooth domes and on the other side we have slowly rising edges
with a flat plateau on top.
A comparison of the finite-size shapes corresponding to different
condensate sizes given in Fig.~\ref{fig:transition-fss}~(a--d)
indicates that the transition line will sharpen with increasing
condensate size, supporting that we are dealing with finite-size
effects. However, the prediction only tells us in the leading order
which phase dominates for infinitely large condensate volumes, but not
whether there is a crossing where both phases are equally likely and
at which volume this would occur. Therefore, the deviation of scaling
exponents from the prediction could also imply that those systems have
higher-order corrections to the transition.

\subsection{Condensate width scaling}

\begin{figure}
  \centering
  \includegraphics{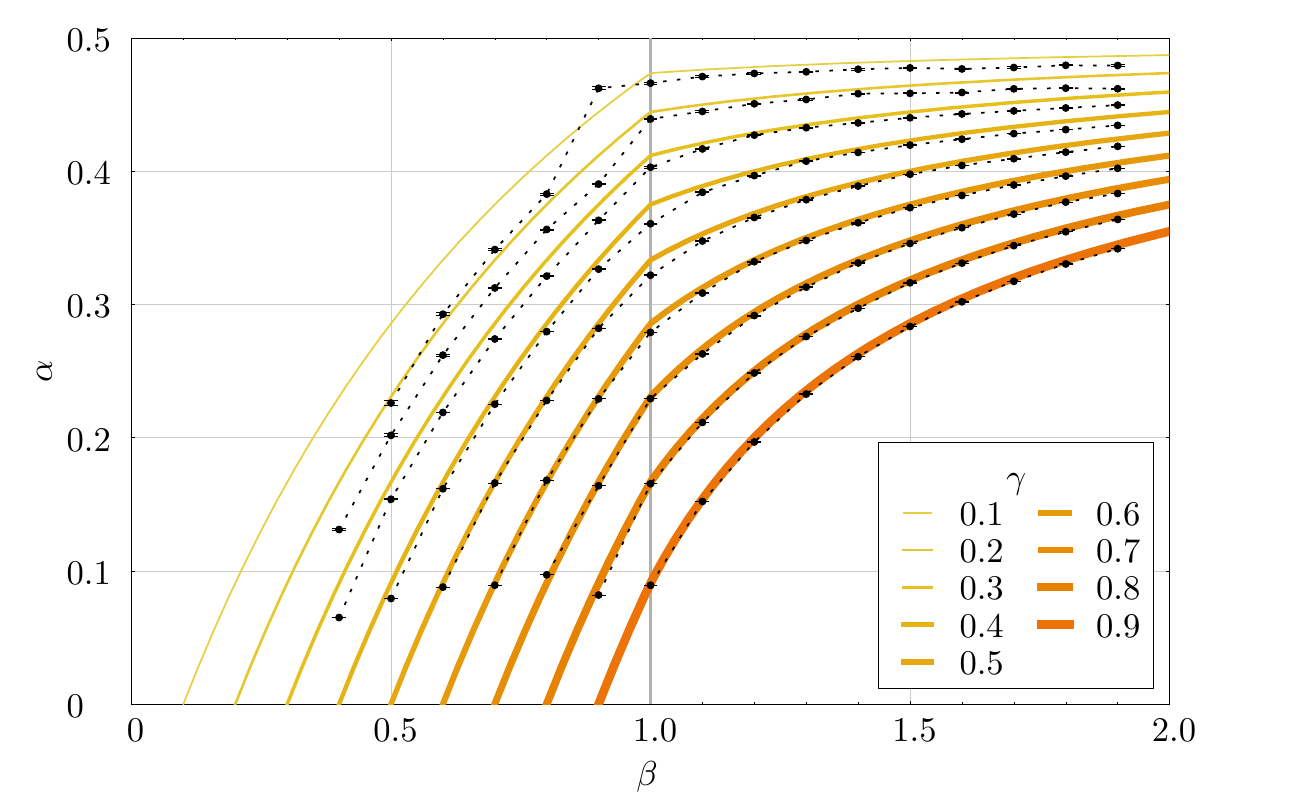}
  \caption{The measured scaling behaviour in the rectangular and
    smooth phases as a function of $\beta$. In contrast to the phase
    diagram in Fig.~\ref {fig:phaseDiagramShapeAdjustingWeights}, the
    scaling exponent $\alpha$ is not shown in a `depth' dimension, but
    as the $y$-value.  Points for the same value of $\gamma$ are
    connected by broken lines.  The continuous colored lines
    correspond to the predicted values for the scaling exponent
    $\alpha$ given by Eq.~(\ref{eq:widthScaling}). The underlying
    condensate volumes are in the range of $3\times 10^3$ to $10^7$,
    depending on how extended the condensates and therefore how
    computationally expensive the simulations are.}
  \label{fig:scalingResults}
\end{figure}

Figure~\ref{fig:scalingResults} shows the estimated scaling exponents
for the width-scaling behaviour of the smooth and rectangular case,
with the measured values listed in Table~\ref{tab:scalingTable}. One
can see that for $\beta>1$ the measurements align well with the
predictions, except for small $\gamma$ values, where the deviations
can be the result of finite-size effects. There, zero-range
interactions are weak and the condensates span many sites, which
eventually becomes unfeasible to simulate for large condensate
volumes. In the case $\beta<1$ and large $\gamma$, the estimates are
very close to the predicted values. Combining the scaling behaviour
with the characteristic shapes, we see that for $\beta > 1.5$ the
condensate shape and scaling exponent is close to the behaviour of
purely interaction-driven condensation, even for $\gamma$ close to~1.

One region where the measurements deviate strongly from the
predictions is for small $\gamma$ in the rectangular case. This region
is also where the scaling behaviour is predicted to change
dramatically at the phase boundary. It looks like the measurements of
scaling exponents change in the same way as the predictions, but
shifted to lower values. We have found no explanation so far why these
systems behave this way. The obvious measurement outlier at position
$\beta=0.9$ and $\gamma=0.1$ can however be explained as an extension
of the smooth scaling behaviour, which agrees with the observation
that the condensate shape has no rectangular characteristics.

The region at $\beta=1$ is of special interest, as there the
condensate not only gradually changes from a smooth to a rectangular
shape as $\gamma$ grows from $0$~to~$1$, but covers the whole scaling
exponent range from $0$~to~$0.5$ as well. In this region it also
becomes feasible to simulate the original dynamics and thus
investigate dynamical properties of these systems as well.

The peak condensate region $\beta<1, \gamma>\beta$, with the
transition line $\beta=\gamma$ is reproduced sufficiently well with
problems at small values of $\gamma$ due to exceedingly large critical
density as visible in the background colour code in
Fig.~\ref{fig:condensateGrid}. 
Here we observe zero-range like condensates as the local interaction
term is strongly dominant.

\begin{table}
  \caption{Table of measured values of scaling exponents $\alpha$ (upper values) and their
    respective absolute deviations from the predicted values (lower values). For the referred
    theoretical predictions see Eq.~\eqref{eq:widthScaling}. To estimate these scaling
    exponents we performed on the order of $10^{9}$ Monte Carlo sweeps to take about $10^{5}$ samples.}
  \centering
  \includegraphics[angle=-90]{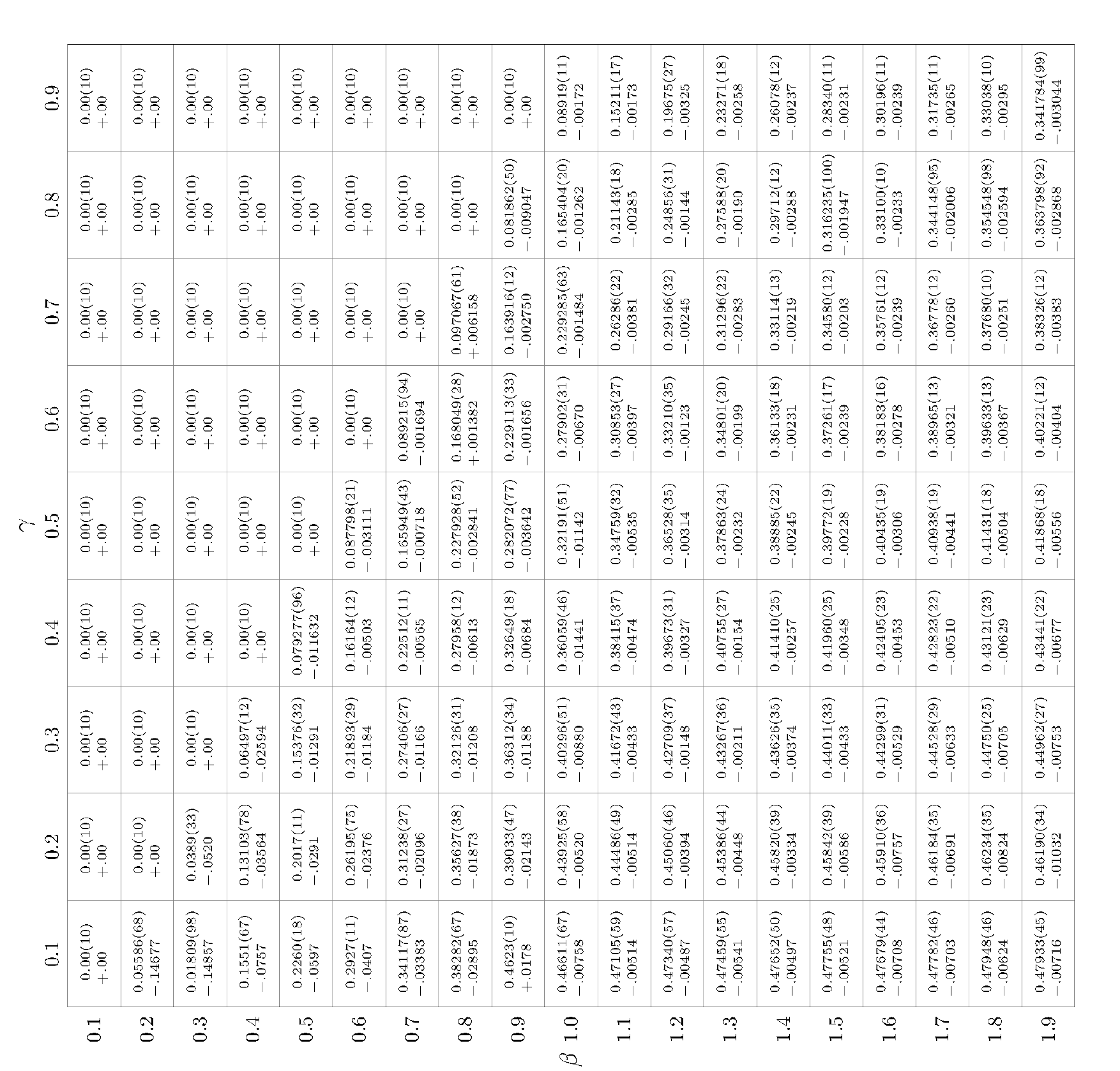}
  \label{tab:scalingTable}
\end{table}

\section{Summary}

Based on the improved simulation method, especially in the rectangular
phase, we were able to show for the pair-factorized steady state
(PFSS) model~\eqref{eq:WaclawWeights} that the three phases predicted
in Refs.~\cite{Waclaw09:tuning-shape,Waclaw09:mass-condensation-1d}
with peak-like, rectangular and smooth (bell-like) condensate shapes
exist in about the regions they are expected in. By analyzing the
characteristic shapes, we found that the finite systems we considered
show finite-size effects that smear out the phase boundary between
rectangular and smooth, bell-like shaped condensates.

With the improved estimator for the condensate scaling behaviour we
were able to confirm that the scaling behaviour is clearly different
from that of purely interaction-driven condensation processes
\cite{Evans06:PFSS}. Table~\ref{tab:scalingTable} demonstrates that
the estimates for large regions of the parameter space are close to
the expected scaling behavior. In the smooth phase, the measured
scaling exponents are within 2\% of the analytically predicted values,
which verifies that the assumptions and approximations in the
analytical predictions of Ref.~\cite{Waclaw09:mass-condensation-1d}
are justified. In fact, the prediction yields very good values of the
scaling exponents even for finite systems provided no transition line
is close. Systematic deviations of the transition line between the
rectangular and smooth condensate shape regimes that we observed are
largely due to finite-size effects.

\ack We would like to thank H.~Meyer-Ortmanns and B.~Wac{\l}aw for
useful discussions and the DFG for financial support under Grant
No.~JA~483/27-1. We further acknowledge support by the DFH-UFA
graduate school under Grant No.~CDFA-02-07.


\end{document}